\documentclass[aps,prb,onecolumn,amsmath,amssymbs,superscriptaddress,reprint,longbibliography]{revtex4-1}
\usepackage{epsfig}
\usepackage{wrapfig}
\usepackage{bbm}
\usepackage[usenames]{color}
\usepackage{array}
\usepackage{times}
\usepackage{braket}
\usepackage{graphicx}
\usepackage{float}
\usepackage{multirow}
\usepackage{natbib,twoopt}
\usepackage[breaklinks=true]{hyperref}
\usepackage{xcolor}
\usepackage{amsmath,amsfonts,amssymb}
\usepackage{graphicx}
\usepackage{hyperref}
\usepackage{color}
\usepackage{stackengine}
\usepackage{subfigure}
\usepackage{verbatim}
\usepackage{stmaryrd} 
\usepackage{lmodern} 
\usepackage{soul}

\newcommand{\RNum}[1]{\uppercase\expandafter{\romannumeral #1\relax}}
\newcommand{\balancecolsandclearpage}{%
	\close@column@grid
	\cleardoublepage
	\twocolumngrid
}


\begin{document}

\title{Coexistence of dipolar and quadrupolar higher-order topology}

\author{Konstantin Rodionenko}
\email{These authors have equally contributed to this work}
\affiliation{School of Physics and Engineering, ITMO University, Saint  Petersburg 197101, Russia}

\author{Maxim Mazanov}
\email{These authors have equally contributed to this work}
\affiliation{School of Physics and Engineering, ITMO University, Saint  Petersburg 197101, Russia}

\author{Maxim A. Gorlach}
\affiliation{School of Physics and Engineering, ITMO University, Saint  Petersburg 197101, Russia}

\begin{abstract}
Two-dimensional higher-order topological insulators are typically classified either as dipolar or quadrupolar depending on the relevant invariant. These two classes were previously considered non-overlapping. Here we put forward an example system exhibiting dipolar and quadrupolar higher-order topology simultaneously, suggest its implementation using the arrays of laser-written evanescently coupled optical waveguides and support our conclusions by the full-wave numerical simulations.
\end{abstract}

\maketitle

\section{Introduction}\label{sec:1}

Higher-order topological insulators (HOTIs), hosting corner and hinge states in two- and three-dimensional systems, have recently been discovered~\cite{Benalcazar_2017_Science,Benalcazar2017,Schindler2018NP} and extensively explored~\cite{Xie_HOTI_review_2021} in photonics~\cite{Mittal_2019,Hassan2019,Kruk2021May,Noh2018Jul}, microwaves~\cite{Peterson2018Mar,Hua_LPR_2020}, electrical circuits~\cite{Imhof2018Sep,Liu2020Aug}, gyromagnetic photonic crystals~\cite{Mele_2020,Zhou2024Oct} and acoustic structures~\cite{Ni2019Feb,Qi2020May,Xue2020May,Xue2019Feb,Hua_PRB_2020,Hua_NC_2020}. 
This extensive interest was largely stimulated by the unique properties of localized topological modes including the robustness of their energy to the disorder ensured by the global properties of the Hamiltonian and crystalline symmetries~\cite{Benalcazar_2017_Science,Quantization_2019}.

In the two-dimensional (2D) systems, the two fundamental types of higher-order topology are the  quadrupole~\cite{Benalcazar2017,Benalcazar_2017_Science} and the dipole~\cite{Quantization_2019} topological insulators. This terminology is rooted in the modern theory of polarization~\cite{Marzari1997Nov,Resta1998Mar,Marzari2012Oct,Benalcazar_2017_Science}, which determines the multipole moments of a periodic solid based on the structure of the Bloch functions.

\begin{figure}[t!]
    \centering
    \includegraphics[width=0.60\linewidth]{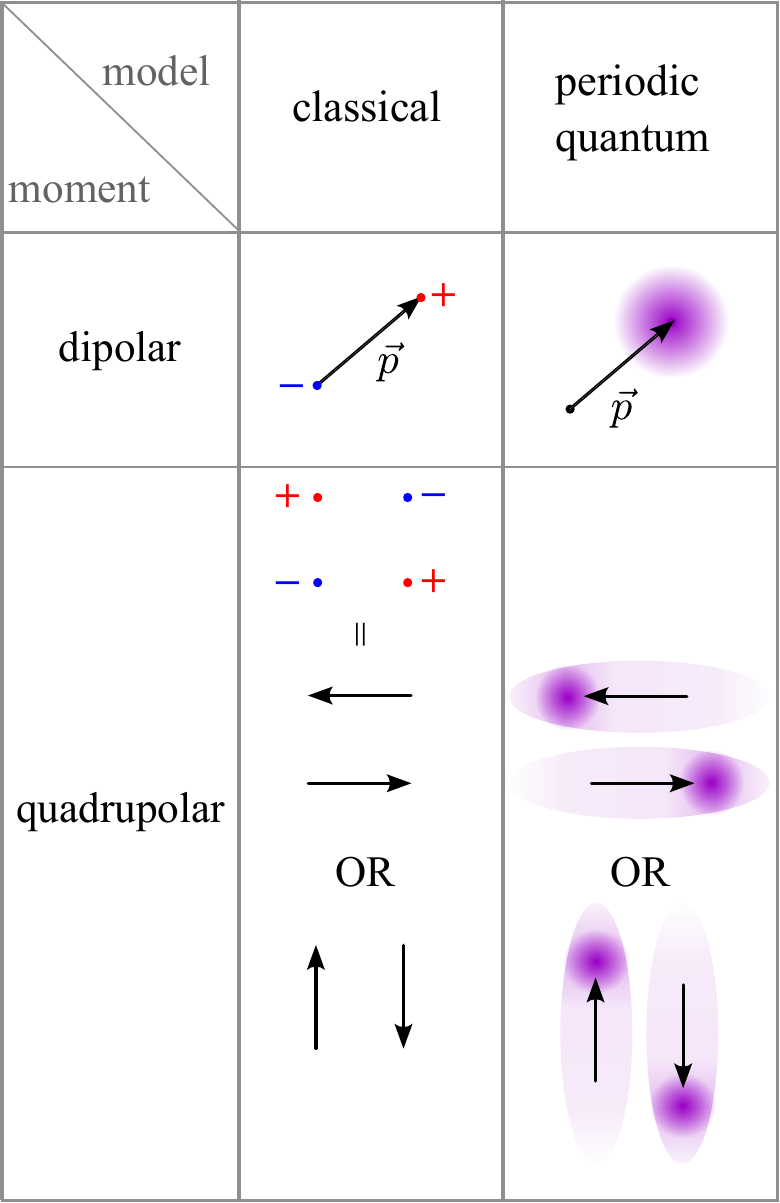}
    \caption{
    Schematic comparison of dipole and quadrupole moments arising in the collection of classical charges (left) and periodic quantum systems (right). 
    }
    \label{Fig0}
\end{figure}

From the classical perspective, any collection of point charges can be assigned a set of the multipole moments, including the net charge, dipole, quadrupole moments and so on. While the total charge typically vanishes due to electroneutrality, all other multipole moments can be simultaneously nonzero and generally depend on the coordinate origin, see Fig.~\ref{Fig0}.

The situation is more complicated in the quantum case. Periodicity of the solid guarantees that the eigenstates can be presented in the form of the Bloch functions. However, the Bloch modes are spatially extended and hence polarization is not well-defined. A solution proposed long ago~\cite{Marzari1997Nov,Resta1998Mar} is to construct the Wannier functions~-- linear combinations of the Bloch modes featuring strong spatial localization. The center-of-mass of those distributions known as the Wannier center directly defines bulk polarization invariant (Fig.~\ref{Fig0}), which is relevant not only for  simple one-dimensional (1D) models~\cite{Zak1989Jun} such as the Su-Schrieffer-Heeger chain~\cite{1DSSH}, but also for the two-dimensional crystalline insulators~\cite{Quantization_2019}.

This solution, however, has a caveat emphasized by Benalcazar and coauthors~\cite{Benalcazar_2017_Science}. In some cases it is fundamentally impossible to construct the Wannier function localized in {\it both} spatial directions simultaneously, since the position operators~\cite{Resta1998Mar} $\hat{x}$ and $\hat{y}$, projected onto the set of the occupied bands, can be non-commuting. This leaves the option to localize the Wannier function in one of the  directions while having it extended in the other. The pair of such semi-localized Wannier functions is analogous to the pair of anti-parallel electric dipoles giving rise to the nonzero quadrupole moment, Fig.~\ref{Fig0}. Computing the expectation value of the projected position operator along the ``delocalized'' direction for a particular Wannier function, or, more generally, particular Wannier subsector~\cite{Hua_PRB_2020,Hua_LPR_2020,Hua_NC_2020}, one extracts the associated dipole moment. 
Notably, such situation occurs in the quadrupole insulators~\cite{Benalcazar_2017_Science,Benalcazar2017}, and the construction of the nested Wilson loops~\cite{Benalcazar2017,Ren2021Jan} essentially reflects the above logic.

Following the pioneering works~\cite{Benalcazar_2017_Science,Benalcazar2017}, 
it is commonly accepted that the bulk quadrupole moment is well-defined only in the systems with the zero bulk dipole polarization and loses meaning otherwise~\cite{Watanabe2020Oct,Ono2019Dec,Vanderbilt2021}. The underlying reason is the origin-shift law: for a chosen set of occupied bands and nonzero bulk polarization, the bulk quadrupole moment is origin-dependent.
Based on that, the families of dipolar and quadrupolar insulators are considered non-overlapping. 

However, this limitation contradicts the classical perspective
(Fig.~\ref{Fig0}) where the charge distribution can feature all kinds of multipole moments simultaneously. Even if the dipole moment is nonzero and the quadrupole moment is origin-dependent, it does not loose its physical meaning and contributes to the field created by the system.

In this Article, we resolve this tension and demonstrate that the higher-order topological insulator (HOTI) can feature nonzero and well-defined dipole and quadrupole moments simultaneously. To support that, we put forward a specific but nontrivial example of a two-dimensional lattice combining both types of the higher-order topology. Furthermore, we develop a systematic approach to compute bulk invariants and demonstrate that their edge and corner signatures are consistent with the bulk-boundary correspondence. 

The rest of the Article is organized as follows. In Sec.~\ref{sec:2} we introduce a model based on the  lattice of multi-mode waveguides featuring coexisting dipolar and quadrupolar topology, construct topological invariants and examine edge and corner states for a finite geometry. Section~\ref{sec:3} extends our analysis further by introducing more fabrication-friendly design based on the single-mode waveguides at the expense of a more complex lattice which can be readily fabricated by the femtosecond laser writing~\cite{szameit_discrete_2005,dipolarFBGraphene,1Ddipole,MultiorbitalVic,Noh2025Dec}. We finally confirm our predictions by the full wave numerical simulations for the designed realistic lattice of optical waveguides.

\begin{figure*}[ht!]
    \centering
    \includegraphics[width=0.90\linewidth]{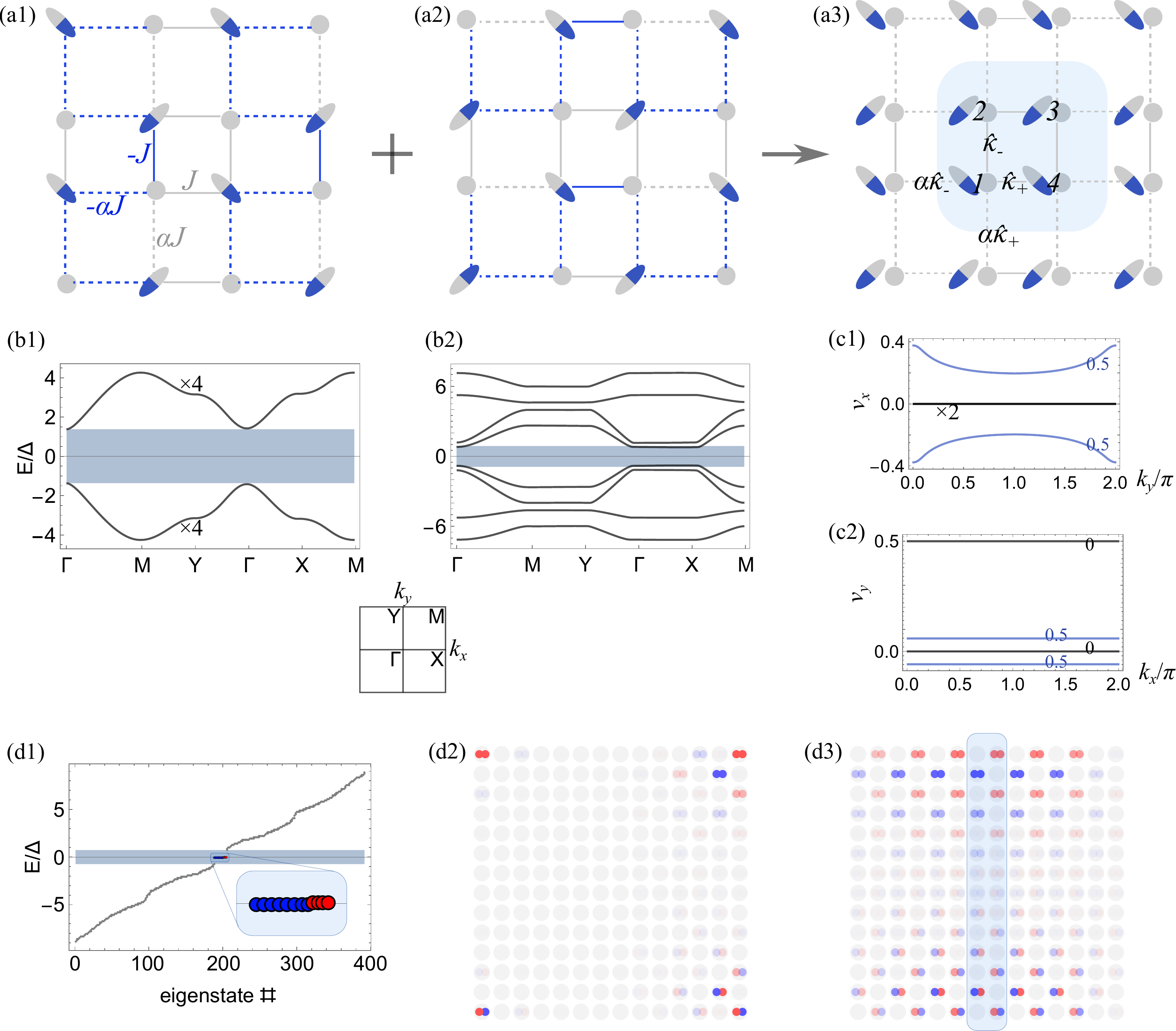}
    \caption{
    (a)~Two-orbital model (a3) obtained by combining two $\pi/2$-rotated copies (a1,a2) of the orbital quadrupole insulator~\cite{Schulz2022Nov}. 
    Positive and negative mode amplitudes and couplings are shown by gray and blue, respectively. The unit cell is shaded.
    (b)~Representative band structures for the doubly-quadrupolar phase with dimerization $\alpha = 2$ and $\kappa = \gamma = 0$ with four-fold degenerate bands (b1) and mixed dipolar-quadrupolar phase at $\kappa = \gamma = \Delta$ (b2). Gray areas highlight the topological bandgaps, while inset shows the first Brillouin zone.
    (c)~Wannier bands $\nu_x(k_y)$ and $\nu_y (k_x)$ obtained via Wilson loops calculation along $k_x$ (c1) and $k_y$ directions (c2). The quadrupolar (dipolar) Wannier subsectors are shown in blue (black). The numbers near each band indicate  Wannier-band polarization; the middle Wannier band in (c1) is doubly degenerate. 
    (d1)~Spectrum of a finite $7 \times 7$ unit-cell lattice in the mixed dipolar-quadrupolar HOTI phase with $\kappa = \gamma = 1.3 \, \Delta$, $\alpha = 2$. 
    The corner states (red) are spectrally separated  from the edge state spectrum (blue) by a small potential $0.02 \Delta$ applied to the four corners. 
    Corner~(d2) and edge~(d3) states amplitude profiles with color-coded phase (red for phase $0$ and blue for phase $\pi$). Amplitudes of $s$ and $p$ modes are schematically represented as left and right colored circles inside the gray lattice sites, respectively. Geometric dimerization is not shown for clarity. Blue region in~(d3) identifies the edge state structure analogous to the end states of the Su-Schrieffer-Heeger model. 
    }
    \label{Fig1}
\end{figure*}

\section{Two-orbital model}\label{sec:2}

As a specific model we consider photonic quadrupole insulator~\cite{Schulz2022Nov}. In contrast to the original system~\cite{Schulz2022Nov}, we assume that each of the waveguides hosts a pair of the degenerate modes (orbitals) with the distinct symmetry of the near field profiles. We demonstrate below that such two-orbital lattice features coexisting dipolar and quadrupolar topological phases.

The canonical four-band model of the quadrupole insulator on a square lattice~\cite{Benalcazar2017,Benalcazar_2017_Science} utilizes a four-site square unit cell with the three positive and one negative coupling amplitudes.

Implementation of this system in the optical range can be expedited~\cite{Schulz2022Nov} by utilizing the waveguides of the two types: one group of the waveguides supports a mode with the full rotational symmetry ($s$-mode), while the second group features an inversion-odd dipole-like mode ($p$-mode). The propagation constants of $s$ and $p$ modes coincide. By properly arranging the waveguides in the lattice [Fig.~\ref{Fig1}(a1)], one can tune the signs of the overlap integrals of the mode fields achieving the desired pattern of the couplings: positive and negative couplings are shown in Fig.~\ref{Fig1}(a1) by gray and blue, respectively. Similarly to the canonical quadrupole insulator~\cite{Benalcazar2017,Benalcazar_2017_Science}, this orbital quadrupole model features 2D geometric dimerization, with inter-cell (intra-cell) coupling being $\pm \alpha J$ ($\pm J$) shown by dashed (solid) lines.

This model supports topologically nontrivial phase with quantized bulk quadrupole moment $q_{xy} = 0.5$ at $\alpha < 1$ which corresponds to the weaker links in the corner unit cells. In this phase, the pair of the Wannier bands $\nu^{x,y}$ constructed from the two lower Bloch bands is gapped, possessing a nested polarization $p^{y,x} = 0.5$. An observable signature of quadrupole topological phase are corner-localized modes arising on all four corners of the lattice.

Now we combine the two copies of the orbital quadrupole lattice  
rotated by $\pi/2$ with respect to each other into a two-orbital model as illustrated in Figs.~\ref{Fig1}(a1-a3). Each site of the lattice now hosts a pair of the degenerate modes ($s$ and $p$) with the distinct inversion symmetry, see Fig.~\ref{Fig1}(a3). By construction, the orientation of the $p$ modes alternates between the rows of the lattice. 

Besides $s$-$p$-type coupling $\Delta$ present in the original orbital quadrupole lattice, our two-orbital model also includes the couplings between the modes of the same symmetry: $s-s$ interaction denoted as $\kappa$ and $p-p$ coupling denoted by $-\gamma$. The minus sign in the latter expression is dictated by the overlap integral between the two $p$ modes.

Assuming perfect degeneracy of $s$ and $p$  modes at each site and using the basis $(s_1,p_1,s_2,p_2,s_3,p_3,s_4,p_4)$ in Fig.~\ref{Fig1}(a3), we recover the Bloch Hamiltonian
\begin{eqnarray}
    \label{Ham1}
    \hat{H} (\textbf{k}) = \left(
        \begin{array}{cccc}
         0 & \hat{\beta}^{-}_{y} & 0 & \hat{\beta}^{+}_{x} \\
         (\hat{\beta}^{-}_{y})^\dagger & 0 & \hat{\beta}^{-}_{x} & 0 \\
         0 & (\hat{\beta}^{-}_{x})^\dagger & 0 & (\hat{\beta}^{-}_{y})^\dagger \\
         (\hat{\beta}^{+}_{x})^\dagger & 0 & \hat{\beta}^{-}_{y} & 0 \\
        \end{array}
    \right)
\end{eqnarray}
where $\hat{\beta}^{\pm}_{x,y} = \hat{\kappa}_\pm + \alpha \hat{\kappa}_\mp e^{- i k_{x,y}}$, and the coupling matrices in the basis of $s$ and $p$ modes read $\hat{\kappa}_{\pm} = 
        \left(
            \begin{array}{cc}
             \kappa & \pm \Delta \\
             \mp \Delta & -\gamma \\
            \end{array}
        \right)
$~\cite{Mazanov2024May,Caceres-Aravena2019Jul,Caceres-Aravena2022Jun}. 
Here, $\kappa$, $-\gamma$ and $\pm \Delta$ are the couplings between similar orbitals $s$ and $s$, $p$ and $p$, and inter-orbital couplings between $s$ and $p$ modes, respectively. Generally, all three types of couplings are comparable in magnitude and together determine the topological phase of the model. For simplicity, we assume that the geometric dimerization of the lattice has a proportional effect on all couplings and is measured by a single parameter $\alpha$ in the intra-cell coupling matrices, $\alpha\,\hat{\kappa}_{\pm}$. 

The model described by the Hamiltonian Eq.~\eqref{Ham1} possesses mirror $M_y$ symmetry such that $ \hat{M}_y \hat{H} (k_x, k_y) \hat{M}_y^{-1} = \hat{H} ( k_x , - k_y )$, chiral symmetry $\hat{\Gamma}_8 = \rm{diag} (1,1, -1,-1, 1,1, -1,-1)$ with $\{ \hat{\Gamma}_8, \hat{H} \} = 0$, and time reversal symmetry $ \hat{\mathcal{T}} \hat{H} (\textbf{k}) \hat{\mathcal{T}}^{-1} = \hat{H} ( - \textbf{k})$. In special cases $\kappa = \gamma = 0$ (uncoupled rotated orbital quadrupole layers) and $\Delta = 0$, an additional $M_x$ symmetry appears. 

The above model hosts three distinct topological phases. First, in the limit of uncoupled quadrupole layers $\kappa = \gamma = 0$, we predictably recover the doubly-degenerate band structure of the single-layer orbital quadrupole lattice~\cite{Schulz2022Nov}, which is further called doubly-quadrupolar phase. The respective band structure is depicted in Fig.~\ref{Fig1}(b1).

As the same-orbital couplings $\kappa$ and $\gamma$ increase, the central bandgap closes at $\Gamma$ point at $\sqrt{\gamma  \kappa } / \Delta = (\alpha -1)/ (\alpha +1)$, and then reopens at $\rm X$ point at $\sqrt{\gamma  \kappa } / \Delta = ( 1 - \alpha^2 + \sqrt{\alpha^4 + 8 \alpha^3 - 2 \alpha^2 + 8 \alpha  + 1} ) / (4 \alpha )$, giving rise to the band structure shown in Fig.~\ref{Fig1}(b2) and a second topological phase. 
As we show below, this case represents a new phase combining dipole and quadrupole higher-order topologies. 

Finally, at even larger $\kappa$ and $\gamma$ so that $\sqrt{\gamma  \kappa } / \Delta = ( \alpha^2 - 1 + \sqrt{\alpha ^4+8 \alpha ^3-2 \alpha ^2+8 \alpha +1} ) / (4 \alpha )$, the bandgap closes again at the $\rm X$ point and the structure enters a metallic phase.

To obtain the bulk invariants in the second topological phase for the representative case  $\kappa=\gamma=\Delta$ [Fig.~\ref{Fig1}(b2)], we calculate the nested Wannier centers via the gauge-consistent nested Wilson loop technique introduced for quadrupolar insulators in Refs.~\cite{Benalcazar_2017_Science,Benalcazar2017,Ren2021Jan}. 
In its first step, similar to dipolar HOTIs~\cite{Quantization_2019}, the Wilson loop $\hat{W}_{k_x}$ ($\hat{W}_{k_y}$) based on four Bloch bands below the bandgap is calculated along $k_x$ ($k_y$) direction, with its eigenvectors defining the linear combinations of these bands localized along $x$ ($y$) axis and extended in the orthogonal direction $y$ ($x$). This procedure provides the semi-localized Wannier functions (see Supplemental Materials for details). 
The corresponding centers of the Wannier functions along $x$ ($y$), i.e. Wannier centers $\nu_{x}(k_y)$ ($\nu_{y}(k_x)$), 
are obtained from the eigenvalues of the Wilson loop, forming four Wannier bands. The calculated bands for the Wilson loop taken in $k_x$ and $k_y$ directions are  shown in Figs.~\ref{Fig1}(c1),(c2), respectively.

We observe that for each direction of the Wilson loop there are Wannier bands with trivial and non-trivial Wannier-band polarization [see black and blue bands in Fig.~\ref{Fig1}(c1,c2)]. Accordingly, we group the Wannier bands into two-band Wannier subsectors shown by blue and black in Fig.~\ref{Fig1}(c1,c2), which are further called quadrupolar and dipolar subsectors.

Given the calculated Wannier bands, we construct the projectors onto the quadrupolar and dipolar subsectors further denoted as $\hat{P}^{\nu_x^{q}}$ and $\hat{P}^{\nu_x^{d}}$ for the Wilson loop along $k_x$ and $\hat{P}^{\nu_y^{q}}$ and $\hat{P}^{\nu_y^{d}}$ for the Wilson loop along $k_y$.

For the quadrupolar Wannier subsector, we find 
non-commutation $[\hat{P}^{\nu_x^{q}} \hat{x} \hat{P}^{\nu_x^{q}}, \hat{P}^{\nu_x^{q}} \hat{y} \hat{P}^{\nu_x^{q}} ] \neq 0$ of the subsector-projected position operators $\hat{P}^{\nu_x^{q}} \hat{x} \hat{P}^{\nu_x^{q}}$ and $\hat{P}^{\nu_x^{q}} \hat{y} \hat{P}^{\nu_x^{q}}$, where the projector onto the quadrupolar Wannier subsector includes the first and the fourth Wannier bands in Fig.~\ref{Fig1}(c1) or first and third Wannier bands in Fig.~\ref{Fig1}(c2) (see Supplementary Materials for details).

In contrast, the position operators projected on the dipolar subsector commute: $[ \hat{P}^{\nu_x^{d}} \hat{x} \hat{P}^{\nu_x^{d}}, \hat{P}^{\nu_x^{d}} \hat{y} \hat{P}^{\nu_x^{d}} ] = 0$, which allows to define 2D Wannier centers for this subsector, as expected for a dipolar HOTI~\cite{Quantization_2019}. 
The net bulk polarization for the dipolar subsector then reads $\mathbf{P} = (0, 0.5)$ and is quantized by  $M_y$ symmetry of the model. 

\begin{figure*}[ht!]
    \centering
    \includegraphics[width=0.97\linewidth]{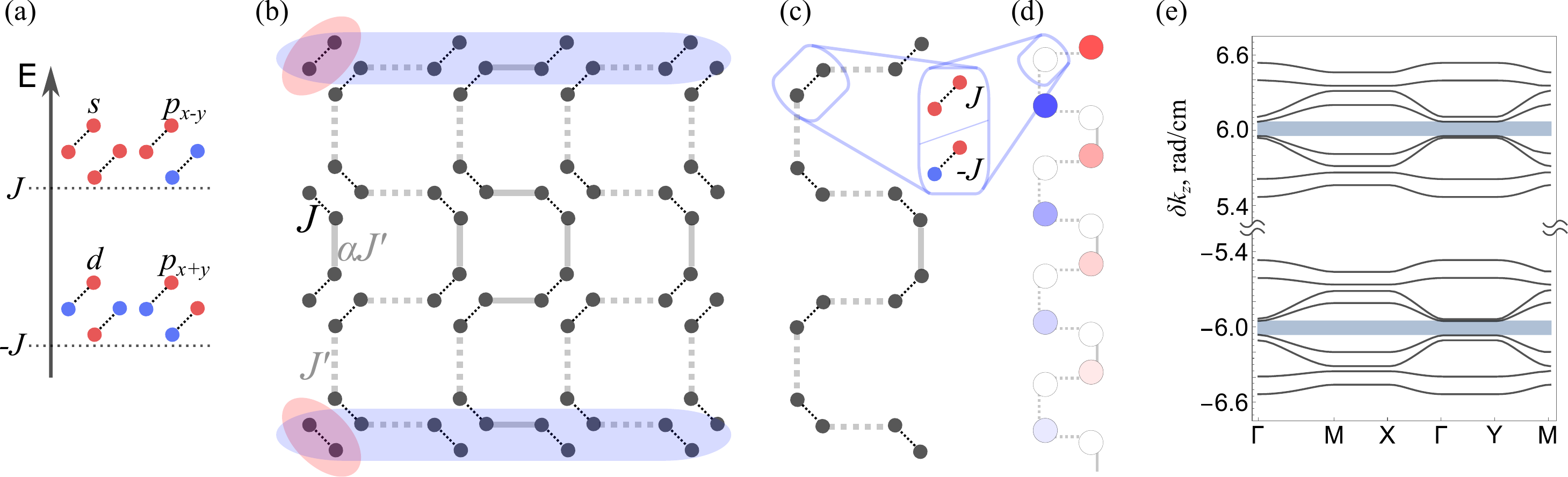}
    \caption{
    (a) The modes of a single four-site meta-atom. Couplings $J$ are shown as dotted links. At energy $-J$, there is a degenerate doublet of quadrupolar ($d$-like) and dipolar ($p_{x+y}$-like) modes, while at $+J$~-- a doublet of monopolar ($s$-like) and orthogonal dipolar ($p_{x-y}$-like) modes. 
    (b) The single-orbital model in the mixed dipolar-quadrupolar HOTI phase near the $\pm J$ energy, where $J', \alpha J' \ll J$. 
    Edge (corner) states are highlighted schematically in blue (red). 
    (c) The model disassembles into uncoupled quasi-1D ``snake''-like generalized SSH chains with alternating nontrivial and trivial topology in the neighboring chains. Each quasi-1D chain contains two practically decoupled spectra formed by anti-symmetric (near the $- J$ energy) and symmetric (near the $+ J$ energy) two-site modes. 
    (d) End states amplitude profile for a single chain in the non-trivial phase ($\alpha = 2$). Here, sites encode dipolar (monopolar) dimer mode amplitudes as a basis for the $d,p_{x+y}$-hybridized ($s,p_{x-y}$-hybridized) sub-models, respectively. 
    (e) Band structure of the single-orbital model for realistic experimental parameters $J=6 \,\, \text{rad}/\text{cm}$, $J'=0.3 \,\, \text{rad}/\text{cm}$, $\alpha = 2$. 
    The two regions of the band structure closely approximate the two-orbital model in the mixed quadrupolar-dipolar HOTI phase [Fig.~\ref{Fig1}(b2)]. 
    Topological bandgaps are highlighted in gray. 
    }
    \label{Fig2}
\end{figure*}

At the same time, the position operators projected onto a {\it single} Wannier band commute. Therefore, for each Wannier band $\nu_{x}(k_y)$ ($\nu_{y}(k_x)$) one could build a set of localized nested Wannier centers via the nested Wilson loop along $k_y$ ($k_x$)~\cite{Benalcazar_2017_Science,Benalcazar2017}. These Wanner centers define the Wannier-band polarizations in Fig.~\ref{Fig1}(c1)-(c2). 
For the bands in the quadrupolar Wannier subsector, we find Wannier-band polarizations $p^{y,x} = 0.5$ with $p^y$ quantized by the $M_y$ symmetry, leading to the bulk quadrupole moment $q_{xy} = 0.5$.

Note that the analogous calculation for the doubly-quadrupolar phase gives two Wannier subsectors both having Wannier-band polarizations $p^{y,x} = 0.5$ and quadrupole moment $q_{xy} = 0.5$  as expected.

This analysis shows that, despite formally not well-defined in the composite model in the presence of nonzero bulk polarization, the bulk quadrupole moment is still well-defined upon separation of the system into the appropriate Wannier subsectors which qualitatively capture the quadrupole and dipole parts of the total polarization.

For the fixed unit cell choice and direction of the Wilson loop, the Wannier centers and nested polarizations (and therefore decomposition into the dipolar and quadrupolar subsystems) are unique and gauge-invariant by construction. At the same time, $k_x$-first and $k_y$-first Wilson loops represent two different splits of the occupied bands, so that the projectors  $\hat{P}^{\nu_x^{q}}$ and $\hat{P}^{\nu_y^{q}}$ do not coincide. Such difference is not surprising and resembles the canonical quadrupole insulator~\cite{Benalcazar2017} where the split into two Wannier bands for $\hat{W}_{k_x}$ and $\hat{W}_{k_y}$ Wilson loops is also different. On a more basic level, this is analogous to the situation when the separation of the spinor wave function onto $\ket{s_z=+1/2}$ and $\ket{s_z=-1/2}$ parts is different from the separation onto $\ket{s_x=+1/2}$ and $\ket{s_x=-1/2}$ contributions.

We also note that $\hat{W}_{k_x}$ and $\hat{W}_{k_y}$ Wilson loops test the signatures of bulk quadrupole moment at the different boundaries, and $q_{xy} = 0.5$ is tied to simultaneously half-quantized nested polarizations $p^{y,x} = 0.5$ in both of the directions~\cite{Yang2023Feb}.

We therefore conclude that the topology of our system  Eq.~\eqref{Ham1} at $\kappa = \gamma = \Delta$ is adiabatically connected to a pair of decoupled 2D quadrupolar and dipolar HOTI layers, even though no spatially homogeneous ($\mathbf{k}$-independent) unitary transformation can split the system directly into quadrupolar and dipolar subsystems.

To reveal the observable signatures of dipole-quadrupole topology, we compute the spectrum and the eigenstates of a finite square lattice including $N\times N$ unit cells. The calculated mode energies are depicted in Fig.~\ref{Fig1}(d1). In agreement with the bulk band structure [Fig.~\ref{Fig1}(b1)], the spectrum features a gap and a set of in-gap modes, localized at the corners and edges of the structure.

Due to the coexistence of bulk dipole and quadrupole higher-order topology, the finite-size lattice hosts both corner and edge states induced by the two distinct mechanisms: bulk quadrupole moment and bulk polarization, respectively.

Nontrivial bulk quadrupole moment $q_{xy} = 0.5$ enables four zero-energy corner states with the profiles shown in Fig.~\ref{Fig1}(d2). Note that the profiles of upper and lower corner states are related by the mirror symmetry $M_y$, while left and right corner modes are inequivalent which is due to the lack of $M_x$ mirror symmetry evident from the lattice structure Fig.~\ref{Fig1}(a3).

In turn, bulk dipole moment $\mathbf{P} = (0, 0.5)$ of the dipolar HOTI Wannier subsector induces $2 (N-1)$ zero-energy edge states on the upper and lower edges of the sample with the typical amplitude profile depicted in Fig.~\ref{Fig1}(d3). Note that the left and right edges of the lattice do not host any edge states since $P_x$ component of the bulk polarization vanishes.

At $\kappa = \gamma = \Delta$ the spatial structure of the two right corner states coincides with the one unit-cell slice of the edge state profile [blue region in Fig.~\ref{Fig1}(d3)], while the left corner states are perfectly localized at a single lattice site. Such decoupling is explained by the destructive interference of same-orbital and inter-orbital couplings and is elaborated for a realistic single-orbital model below.

For other parameters in the dipole-quadrupole phase, such as $\kappa = \gamma = 1.3\, \Delta$ in Fig.~\ref{Fig1}(d1)-(d3), the corner and edge states penetrate exponentially into the bulk of the lattice. 

Finally, as we show in the Supplementary Materials, energies of both corner and edge states are robust to the disorder both in the couplings and on-site symmetric orbital detuning. In this sense, the topological protection of the structure resembles that in the canonical dipolar and quadrupolar HOTIs.

\section{Single-orbital model}\label{sec:3}

While multi-mode $s-p$ meta-atoms require numerical fine-tuning and may be challenging to fabricate, a similar dipolar-quadrupolar HOTI phase can be achieved in a more experimentally feasible setting. As a building block of the lattice we use a rhombic four-site meta-atom with two pair-wise couplings $J$ acting effectively as a two-mode element near the energies $\pm J$ [Fig.~\ref{Fig2}(a)]. 

In a lattice, the degeneracy of the meta-atom modes at $\pm J$ is lifted by the couplings $|J'| \ll J$ [Fig.~\ref{Fig2}(b)] giving rise to the 
band structure with nontrivial gaps near $\pm J$ corresponding to the desired mixed HOTI phase. 

We support this prediction by demonstrating both corner end edge states in the full-wave simulations of the optical $p$-mode waveguide lattice, which can be readily fabricated by the conventional femtosecond laser writing technique.

The meta-atom under study possesses two sets of symmetric degenerate modes [Fig.~\ref{Fig2}(a)]: a quadrupolar ($d$-like) and dipolar ($p_{x+y}$-like) modes at $-J$, and a monopolar ($s$-like) and orthogonal dipolar ($p_{x-y}$-like) modes at $+J$. When arranged into a geometrically dimerized lattice with coupling strengths $J', \alpha J' \ll J$ [Fig.~\ref{Fig2}(b), two band structures form around the energies $\pm J$. These bands are well approximated by the two-band model with $s,p_{x-y}$ and $d,p_{x+y}$ degeneracies, respectively. Note that the condition $\kappa = \gamma = \Delta = J'/4$ is ensured automatically due to the symmetry of the rhombus meta-atom modes.

\begin{figure*}[ht!]
    \centering
    \includegraphics[width=0.99\linewidth]{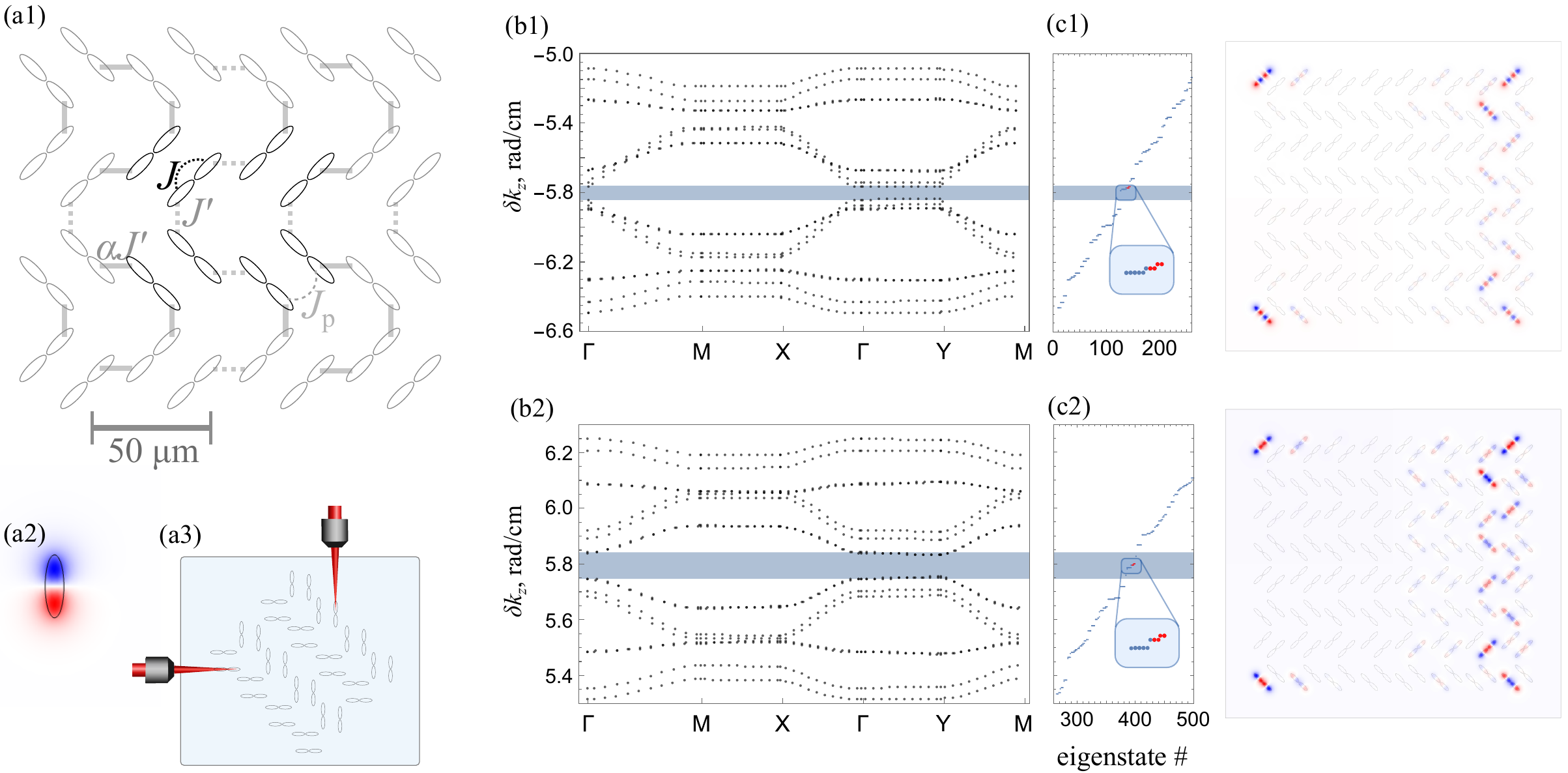}
    \caption{
    (a1) Geometry of the unit cell for the optical waveguide lattice in full-wave COMSOL simulation. Waveguides are assumed to be elliptical with refractive index contrast $\delta n = 0.002$, semi-axes $2.2 \,\mu$m and $7.4 \,\mu$m. Ambient refractive index is set to $n = 1.48$ corresponding to borosilicate glass. 
    (a2) The $p$-mode profile for the isolated waveguide. 
    (a3) Possible sequential fabrication method via the femtosecond laser writing technique from the two facets of the glass wafer. 
    (b) Numerically calculated lower~(b1) and upper~(b2) band structures. 
    (c1,c2) Corresponding numerical spectrum for a finite $4 \times 4$-unit-cell lattice, with localized corner states amplitude profiles shown on the right. 
    }
    \label{Fig3}
\end{figure*}

By construction, the model in Fig.~\ref{Fig2}(b) near $\pm J$ can be alternatively presented as a collection of uncoupled ``snake''-like generalized Su-Schrieffer-Heeger~\cite{1DSSH} (SSH) lattices with dimer meta-atoms highlighted in Fig.~\ref{Fig2}(c). %
When unfolded, 
the chains are described by simple 1D Hamiltonians 
\begin{eqnarray}
    \hat{H}_{\text{chain}} = 
    \left(
        \begin{array}{cccc}
         0 & \tilde{J}' & 0 & \alpha  J' e^{-i k_x} \\
         \tilde{J}' & 0 & \tilde{J}' & 0 \\
         0 & \tilde{J}' & 0 & \tilde{J}' \\
         \alpha  \tilde{J}' e^{i k_x} & 0 & \tilde{J}' & 0 \\
        \end{array}
    \right), 
\end{eqnarray}
where $\tilde{J}' = J' / \sqrt{2}$ is the effective dimer coupling. 
The chain Hamiltonian has a band structure with three gaps and the middle (zero-energy) gap has nontrivial or trivial Zak phase~\cite{Zak1989Jun} depending on whether $\alpha > 1$ or $\alpha < 1$. 
Thus, the neighboring chains in the original lattice [Fig.~\ref{Fig2}(b)] alternate between nontrivial and trivial phases.

Returning to the two-dimensional lattice, the edge and corner states can then be interpreted as stemming from the end states of the uncoupled generalized SSH chains. 
The SSH-like structure of the end states in the chains in the non-trivial phase shown in Fig.~\ref{Fig2}(d) mirrors the edge and right corner state structure highlighted in blue in Fig.~\ref{Fig1}(d3). The isolated left corner states, on the other hand, correspond to the dimers decoupled from the rest of the lattice by construction, as highlighted by red in Fig.~\ref{Fig2}(b). 

Flat regions in the band structure of the original two-mode model [Fig.~\ref{Fig1}(b2)] correspond to the decoupled chains in horizontal directions of the Brillouin zone $M \rightarrow Y$, $\Gamma \rightarrow X$, while dispersive regions capture the dispersion of excitations propagating along the chain.

Finally, moderate coupling of the dimers in the rhombus leads to the coupling of the generalized SSH chains, while in the original two-orbital model it corresponds to shifting from the condition $\kappa = \gamma = \Delta$ and lifting of degeneracies. This in turn leads to exponential, rather than complete, localization of the corner modes in the horizontal direction, consistent with Fig.~\ref{Fig1}(d2)-(d3).

We realize the system numerically on the basis of optical waveguide platform at working wavelength $\lambda = 730$~nm with lattice geometry shown in Fig.~\ref{Fig3}(a1). 
Such elliptic waveguide lattice could be fabricated via the conventional femtosecond laser writing techniques~\cite{szameit_discrete_2005,dipolarFBGraphene,1Ddipole,MultiorbitalVic,Noh2025Dec} in two sequential steps of writing from the two facets of the glass wafer, see Fig.~\ref{Fig3}(a3). 
Alternatively, one could use micro-printing setups~\cite{Schulz2021Aug} such as one used in recent optical realization of the quadrupolar HOTI induced by orbital flux~\cite{Schulz2022Nov}. 

In order to predictably arrive at the mixed quadrupolar-dipolar phase as described in the previous section, we fine-tune the couplings 
to the values as in Fig.~\ref{Fig2} 
by changing the corresponding distances (see geometric and fine-tuning details in Supplemental Materials). 
First, we set the coupling within the waveguide dimers to $J = 6$~rad/cm. 
Next, we set the coupling $J_p$ between the two pairs of waveguides in the meta-atom to a negligible level $J_p = 0.06 \text{ rad/cm } \ll J'$. 
Finally, we arrange equal horizontal and vertical couplings inside and between the the unit cells, realizing $J' = 0.3$~rad/cm and $\alpha = 2$. 
For the finite lattice, we also tune the small on-site potential at the four corners to be $\delta = 0.005$~rad/cm~$\sim 0.017 J'$ by adding the extra refractive index contrast $\delta n' \sim 9 \cdot 10^{-8}$ at the corner rhombic meta-atoms. This additional potential is only intended to spectrally separate the corner states from the edge state band.

The band structure of the periodic lattice is shown in Fig.~\ref{Fig3}(b1-b2). 
Note that the additional approximate $E_x$-$E_y$ polarization degeneracy appears as a feature of elliptical optical waveguides. 
The bandgaps are of nearly the same size for the lower~(b1) and upper~(b2) band structures, although dispersion appears somewhat different. 
This is caused mainly by the limitations of the tight-binding model with effective coupling $J$ to accurately describe the interaction of two closely spaced waveguides. Nevertheless, the band structures and bandgaps are nontrivial, as it is evidenced by the the corner and edge states in the numerical finite lattice spectrum shown in Fig.~\ref{Fig3}(c1-c2).

From the practical perspective, the main requirement for realizing mixed dipolar-quadrupolar HOTI phase is for the individual waveguides to support elongated modes so that the coupling $J_p$ is negligible, while exact waveguide profiles could vary. The remaining fine-tuning is achieved by choosing appropriate geometric anisotropy of the lattice spacings.

\section{Discussion}\label{sec:Discussion}

In summary, we have shown that the two types of higher-order topology, namely, dipolar and quadrupolar, can coexist in a single structure and can be evaluated independently in the Wannier sector representation. As we prove, depending on the chosen set of the Wannier bands, projected position operators are either commuting or non-commuting, signifying dipolar or quadrupolar higher-order topology, respectively. This observation enriches the variety of topological phases allowing to combine different multipole topologies similarly to the different multipoles arising in the field expansion of a classical charge distribution.

As a result, the edge and corner states in our structure are controlled by the different types of topological invariants, which could lead to the different degree of topological protection of these two classes of modes.

An interesting future direction is the bulk signatures of mixed dipole-quadrupole topology. From one hand, nontrivial quadrupole topology is known to produce non-Abelian Berry curvature, resulting in the non-Abelian Bloch oscillations~\cite{DiLiberto2020Nov}. At the same time, these effects disappear for the usual Wannier-type HOTIs. Given that the dipole and the quadrupole Wannier subsectors coexist in the same energy window, this means that the dipole-quadrupole insulator can feature non-Abelian Bloch oscillations, but only for a particular class of the initial conditions, which is an interesting topic for future studies.

\section*{Acknowledgments}
This work was supported by the Russian Science Foundation grant No.~25-79-31027.

\bibliography{refs}

\end{document}


\title{Supplemental Materials: \\ Coexistence of dipolar and quadrupolar higher-order topology}

\author{Konstantin Rodionenko}
\email{These authors have equally contributed to this work}
\affiliation{School of Physics and Engineering, ITMO University, Saint  Petersburg 197101, Russia}

\author{Maxim Mazanov}
\email{These authors have equally contributed to this work}
\affiliation{School of Physics and Engineering, ITMO University, Saint  Petersburg 197101, Russia}

\author{Maxim A. Gorlach}
\affiliation{School of Physics and Engineering, ITMO University, Saint  Petersburg 197101, Russia}

\maketitle

\onecolumngrid

\setcounter{equation}{0}
\setcounter{figure}{0}
\setcounter{table}{0}
\setcounter{page}{1}
\setcounter{section}{0}
\makeatletter
\renewcommand{\theequation}{S\arabic{equation}}
\renewcommand{\thefigure}{S\arabic{figure}}
\renewcommand{\bibnumfmt}[1]{[S#1]}
\renewcommand{\citenumfont}[1]{S#1}

\tableofcontents

\section{1. Wilson loop calculation}

We obtain the Wannier bands and nested Wanner polarizations, as well as position operators projected onto the Wannier subsectors, by the Wilson loop and nested Wilson loop calculations~\cite{Benalcazar_2017_PRB,Benalcazar_2017_Science} detailed below. 

The Wilson loop along $k_x$ ($k_y$) allows to calculate the Wannier centers $\nu_x (k_y)$ ($\nu_y (k_x)$)~-- eigenvalues of position operators projected onto the bands below the bandgap (in our case, four bands). For this, the position operators are discretized in the Brilloin zone. 
The projected position operator $\hat{P}\hat{x}\hat{P}$ has a simple form in the basis of Bloch eigenstates $\gamma_{n,\mathbf{k}} | 0 \rangle$. We order the columns consistently with the calculation of the Wilson loops by $k_x \in \{k_{x1}, k_{x2}=k_{x1} +(2\pi/N)), ... k_{xN}=k_{x1} + (2\pi(N-1)/N))\}$ for each given $k_y$, with $N$ the number of points in the Wilson line, i.e. $ 
(k_{x1},k_{y1}), (k_{x2},k_{y1}), ... , (k_{xN},k_{y1}); 
(k_{x1},$
$
k_{y2}), (k_{x2},k_{y2}), ... , (k_{xN},k_{y2});
...; 
(k_{x1},k_{yN}), (k_{x2},k_{yN}), ... , 
$ $
(k_{xN},k_{yN})
$. 
In this basis, $\hat{P}\hat{x}\hat{P}$ has the block-diagonal form~\cite{Benalcazar_2017_PRB,Benalcazar_2017_Science} 
\begin{eqnarray}
\label{PxP1}
    \hat{P}\hat{x}\hat{P}
    =
    \left(\begin{array}{ccccc}
    \hat{W}_{k_{y1}} & 0 & 0 & \ldots & 0 \\
    0 & \hat{W}_{k_{y2}} & 0 & \ldots & 0 \\
    0 & 0 & \hat{W}_{k_{y3}} & \ldots & 0 \\
    \vdots & \vdots & \vdots & \ddots & \vdots \\
    0 & 0 & 0 & \ldots & \hat{W}_{k_{yN}}
    \end{array}\right)
\end{eqnarray}
with
\begin{eqnarray}
\label{PxP2}
    \hat{W}_{k_{yi}}
    =
    \left(\begin{array}{ccccc}
    0 & 0 & 0 & \ldots & \hat{F}_{k_{xN},k_{yi}} \\
    \hat{F}_{k_{x1},k_{yi}} & 0 & 0 & \ldots & 0 \\
    0 & \hat{F}_{k_{x2},k_{yi}} & 0 & \ldots & 0 \\
    \vdots & \vdots & \vdots & \ddots & \vdots \\
    0 & 0 & 0 & \ldots & 0
    \end{array}\right)
    . 
\end{eqnarray}
Here, $\hat{F}_{k_{xN},k_{yi}} = \hat{U} \hat{V}^\dag$ are constructed from the singular-value decomposition $ \hat{G} = \hat{U} \hat{D} \hat{V}^\dag $ of the Wilson line element $\left[G_{k_{xj},k_{yi}} \right]^{m n}=\left\langle u_{k_{xj}+\Delta_k, k_{yi}}^n \mid u_{k_{xj}, k_{yi}}^m\right\rangle$, where upper indices mark the occupied bands and $\Delta k = 2 \pi / N$. 
The projected position operator $\hat{P}\hat{y}\hat{P}$ is constructed analogously. Note that in order to compute the commutator, both position operators should be transformed to the same basis by a corresponding permutation.

The position operators projected onto the particular two-Wannier-band subsectors $\nu_{x,y}$, $\hat{P}^{\nu_{x,y}}\hat{x}\hat{P}^{\nu_{x,y}}$ and $\hat{P}^{\nu_{x,y}}\hat{y}\hat{P}^{\nu_{x,y}}$~\cite{Benalcazar_2017_PRB,Benalcazar_2017_Science} could be constructed analogously. 
Their eigenvalues are the nested Wilson loop eigenvalues~\cite{Benalcazar_2017_PRB,Benalcazar_2017_Science} (nested Wannier centers) and define the topological invariant for the quadrupolar subsystem, while their commutator distinguishes the quadrupolar and dipolar types of higher-order topology. 
The structure of such projected position operators is identical to Eqs.~\eqref{PxP1}-\eqref{PxP2}~\cite{Benalcazar_2017_PRB,Benalcazar_2017_Science}, with elements 
$F^{\nu_a}_{k_{xN},k_{yi}} = \hat{U} \hat{V}^\dag$ constructed from the singular-value decomposition $ \hat{G}^{\nu_{a}} = \hat{U} \hat{D} \hat{V}^\dag $
of the Wilson line element $\left[G^{\nu_{a}}_{k_{x},k_{y}} \right]^{n n'}
=
[\nu^j_{a,(k_x,k_y+\Delta k_y)}]^n
\langle w^j_{a,(k_x,k_y+\Delta k_y)} | w^j_{a,(k_x,k_y)} \rangle
[\nu^{j*}_{a,(k_x,k_y)}]^{n'}
$, where $a=x,y$ is the direction of the associated Wilson line, $j$ marks the number of the considered Wannier subsector, $[\nu^j_{a,(k_x,k_y)}]^n$ is the $n^\text{th}$ component of the corresponding Wannier eigenstate, and $|w^j_{a,(k_x,k_y)} \rangle$ is the corresponding Wannier function constructed from the Bloch eigenvector $u(k_x,k_y)$ as $|w^j_{a,(k_x,k_y)} \rangle = \sum_{n} [\nu^{j}_{a,(k_x,k_y)}]^{n} [u(k_x,k_y)]^{n}$.

\section{2. Numerical fine-tuning of lattice geometry}

\begin{figure*}[ht!]
    \centering
    \includegraphics[width=0.50\linewidth]{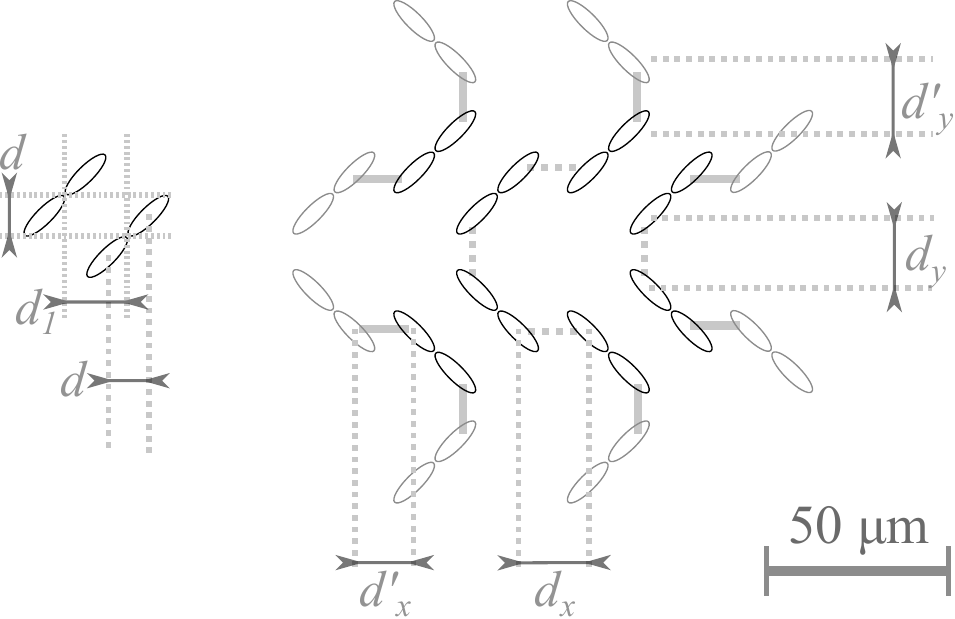}
    \caption{
    Left panel: geometry of the meta-atom. 
    Right panel: geometry of the unit cell. 
    }
    \label{Fig_geometry}
\end{figure*}

\begin{figure*}[ht!]
    \centering
    \includegraphics[width=0.75\linewidth]{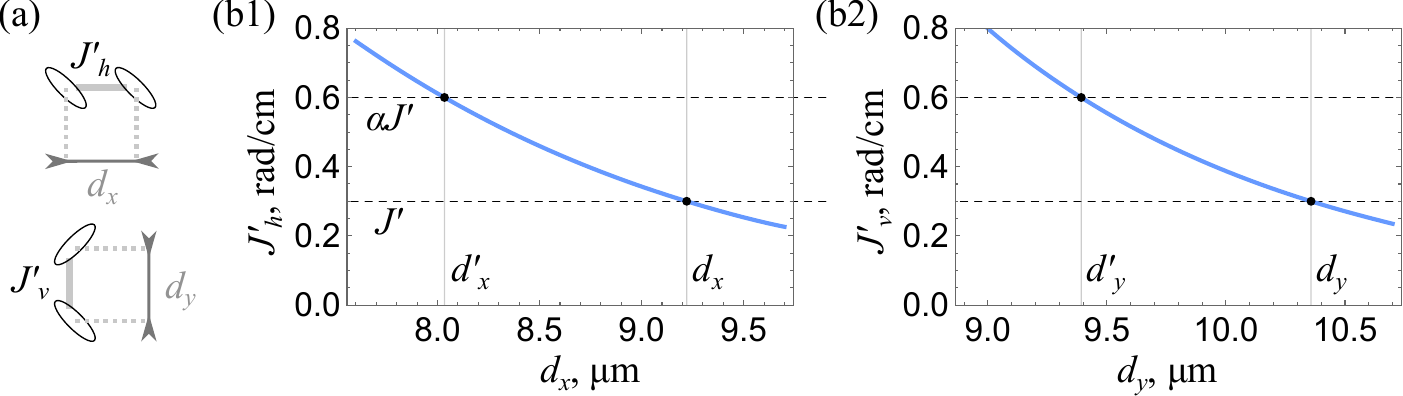}
    \caption{
    (a) Geometry of the horizontal and vertical couplings appearing in the optical waveguide model. (b1) Coupling $J_{h}$ as function of distance $d_x$. (b2) Coupling $J_{v}$ as function of distance $d_y$. 
    The horizontal dashed lines on both plots corresponds to the choices of distances $d_x$, $d_x'$, $d_y$, $d_y'$ at which the respective couplings $J_h (d_x) = J_v (d_y) = 0.3 \,\, \text{rad}/\text{cm}$, $J_h (d_x') = J_v (d_y') = \alpha J_h (d_x) = \alpha J_v (d_y) = 0.6 \,\, \text{rad}/\text{cm}$ where $\alpha = 2$, and which are further used in the full two-dimensional lattice geometry. 
    }
    \label{FigS_couplings}
\end{figure*}

The geometry of the numerical model is shown in Fig.~\ref{Fig_geometry}. Here, $d = 11 \,\mu\text{m}$, $d_1 = 17 \,\mu\text{m}$, $d_x = 18.6 \,\mu\text{m}$, $d'_x = 16.14 \,\mu\text{m}$, $d_y = 20.72 \,\mu\text{m}$, $d'_y = 18.78 \,\mu\text{m}$. 
Having fixed the base geometric parameter $d = 11 \,\mu\text{m}$ defining the relative positions of waveguides in a pair with largest coupling $J$ as well as the vertical shift between the pairs in the rhombic meta-atom, we arrive to such geometric parameters via a fine-tuning procedure, requiring that $J_x = J_y \equiv J$, $J'_x = J'_y \equiv \alpha J$ with $\alpha = 2$, while keeping negligible coupling $J_p \ll J',J$. The corresponding coupling curves are presented in Fig.~\ref{FigS_couplings}.

\begin{figure*}[ht!]
    \centering
    \includegraphics[width=0.80\linewidth]{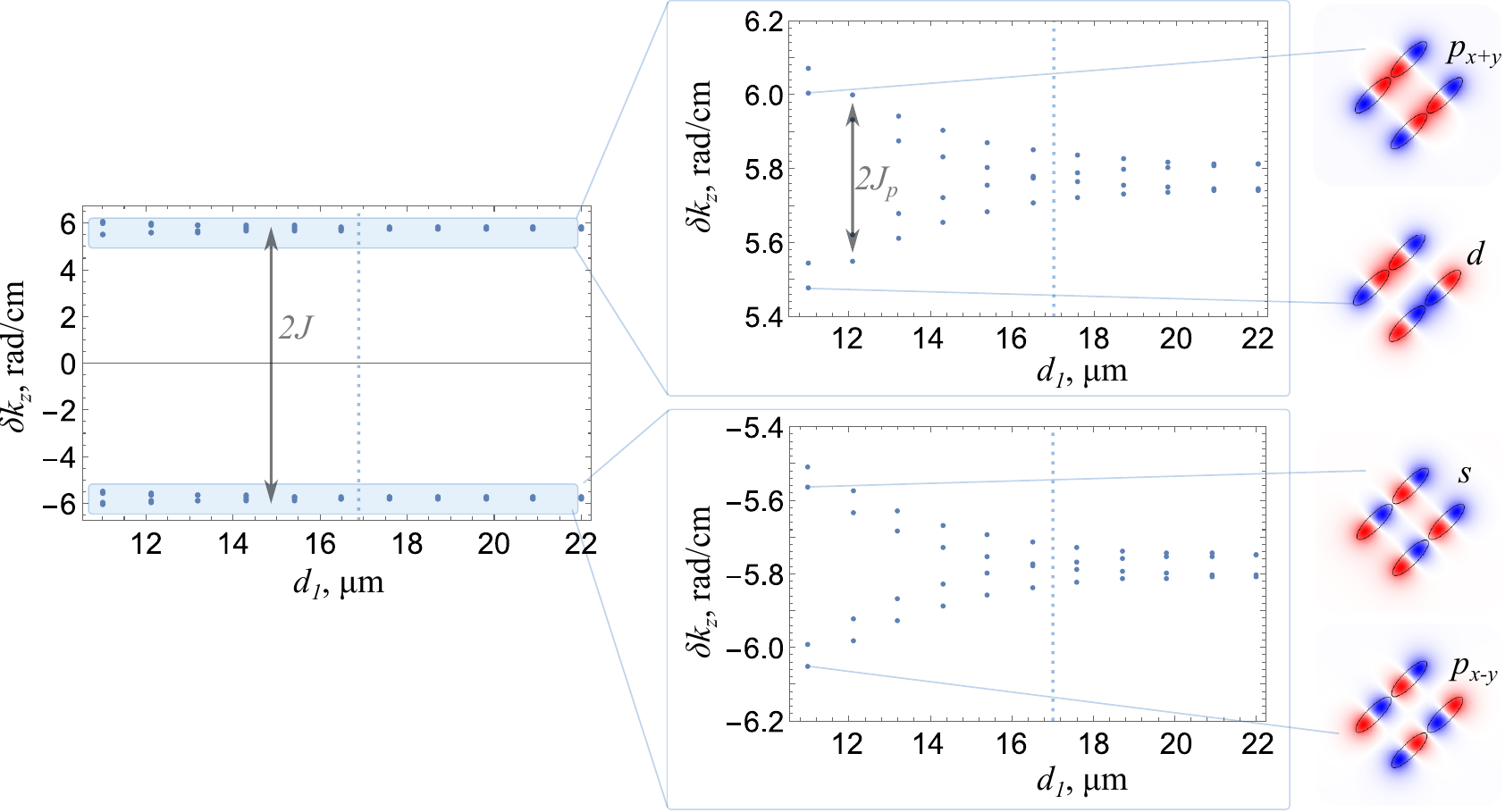}
    \caption{
    Spectrum of the rhombic optical waveguide meta-atom as function of $d_1$ for fixed $d = 18.6 \,\mu\text{m}$. 
    Right panel shows the two enlarged regions of the spectrum and $E_x$ field profiles for the corresponding modes with polarization orthogonal to the waveguides (remaining four nearly-degenerate modes with polarization along the waveguides have analogous profiles). 
    Dashed vertical line corresponds to the chosen value of $d_1 = 8.5 \,\mu\text{m}$ where the coupling $J_p$ becomes negligible (of the order of polarizational degeneracy imperfection). 
    }
    \label{VS_d1}
\end{figure*}

We also tune the near-zero coupling $J_p$ by controlling the horizontal shift $d_1$, see Fig.~\ref{VS_d1}. 
Note that the order of states in the spectrum reverses compared to Fig.2 in the main text since the the coupling $J$ between the rhombus ``edges'' formed by waveguide pairs has negative sign for the considered $p$-modes of individual waveguides. As discussed in the main text, the double amount of states comes from the near-degeneracy in polarization.

\section{3. Dependence of Wannier bands and nested polarizations on the choice of unit cell}

\begin{figure*}[h!]
    \centering
    \includegraphics[width=0.45\linewidth]{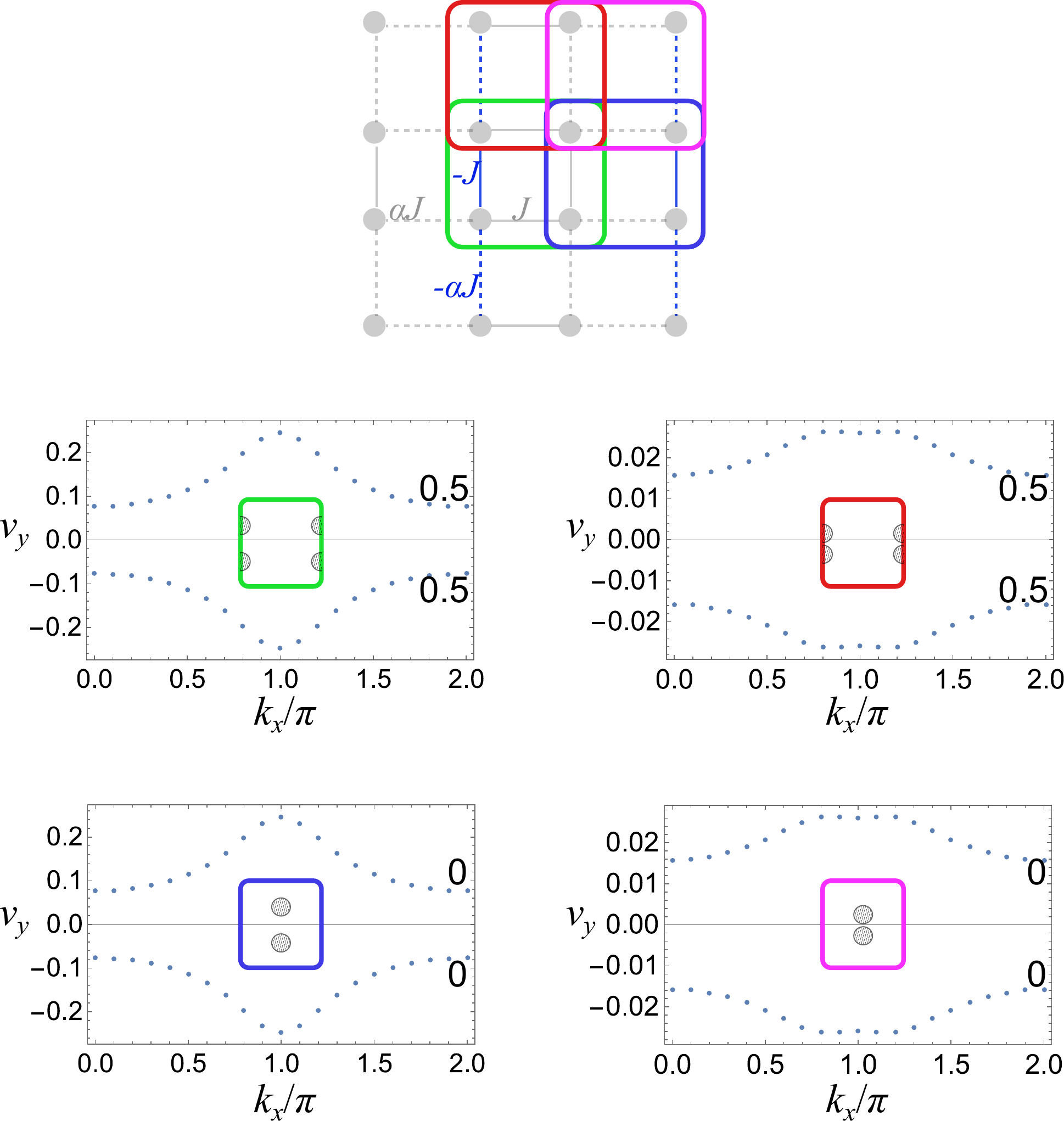}
    \caption{
    Wannier bands and nested Wannier polarizations (indicated near corresponding Wannier bands) for the four variants of unit cell choice, for the canonical quadrupole model in the topologically nontrivial phase with $\alpha = 2$. The circles mark the positions of nested Wannier centers. 
    }
    \label{change_basis_quadrupole}
\end{figure*}

\begin{figure*}[h!]
    \centering
    \includegraphics[width=0.55\linewidth]{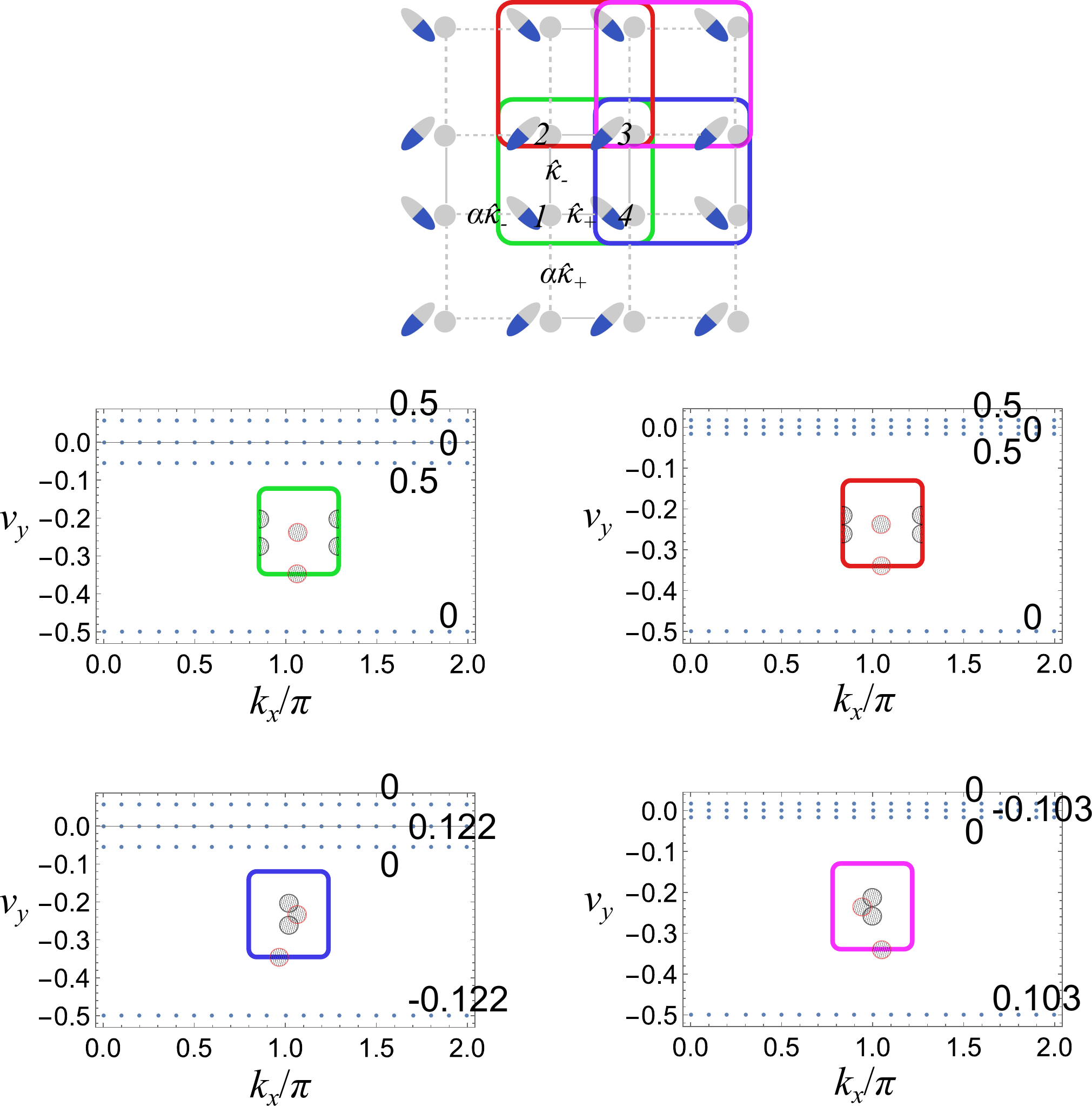}
    \caption{
    Same quantities as in Fig.~\ref{change_basis_quadrupole} for the mixed dipolar-quadrupolar phase in the two-mode model with $\kappa = \gamma = \Delta$, $\alpha = 2$. 
    }
    \label{change_basis_dquadrupole}
\end{figure*}

Figs.~\ref{change_basis_quadrupole}-\ref{change_basis_dquadrupole} show the Wannier bands and nested Wannier polarizations for the four choices of the unit cell in the single-orbital canonical quadrupole insulator model~\cite{Benalcazar_2017_PRB,Benalcazar_2017_Science} [Fig.~\ref{change_basis_quadrupole}] and the two-orbital mixed dipole-quadrupole phase of the model discussed in the main text [Fig.~\ref{change_basis_dquadrupole}]. 

Notably, not only the Wannier centers change with the change of the unit cell choice for their calculation, but also the nested polarizations change as well. 

As an example, we take the first Wilson loop directed along $k_x$. 
In the canonical single-orbital model of quadrupole insulator, for this choice of the Wilson loop, shift of unit cell upwards by half-period does not change the nested polarizations and the quadrupole moment, while shift by half-period to the right trivializes both, see Figs.~\ref{change_basis_quadrupole}. Intuitively, this trivialization can be understood by the shift of position of nested Wannier centers (shown schematically by the circles) by a half-period to the right. 
The calculated projected position operators commutator is nonzero for all four unit-cell choices.

In the mixed quadrupolar-dipolar phase of the two-mode model, such redefinition of unit cell results in even more stark changes, Figs.~\ref{change_basis_dquadrupole}. Indeed, for the shift upwards, the qualitative picture stays the same and the quadrupolar-dipolar decomposition based on Wannier centers holds. However, for the shift to the right or combined right-upward shift, much like in the single-orbital quadrupolar case, the previously quadrupolar sector trivializes; the dipolar one, at the same time, acquires nonzero nested polarizations. 
The calculated projected position operators commutator is zero for all Wannier bands with zero nested polarization, and nonzero otherwise. 
Hence, the dipolar subsystem fully absorbs the quadrupolar characteristics for the right- and right-upward-shifted choice of the unit cell.

Thus, we conclude that in both single-orbital quadrupolar and two-orbital quadrupolar-dipolar cases, the Wannier bands and nested polarizations, and, therefore, the quadrupole moment, depend on the choice of the unit cell. 

This echoes the situation in the celebrated Su-Schrieffer-Heeger model where depending on the unit cell choice the Zak phase is either zero or $\pi$, indicating that the topological edge state arises only for a specific lattice termination. In a similar, but more complex, way the choice of the unit cell affects the topological characteristics of our model.

Nevetheless, there exists a well-defined choice of the unit cell and the lattice termination which allows to clearly separate the dipole and the quadrupole parts of bulk polarization.

%

\section{4. Disorder robustness of the zero-energy modes}

\begin{figure*}[ht!]
    \centering
    \includegraphics[width=0.60\linewidth]{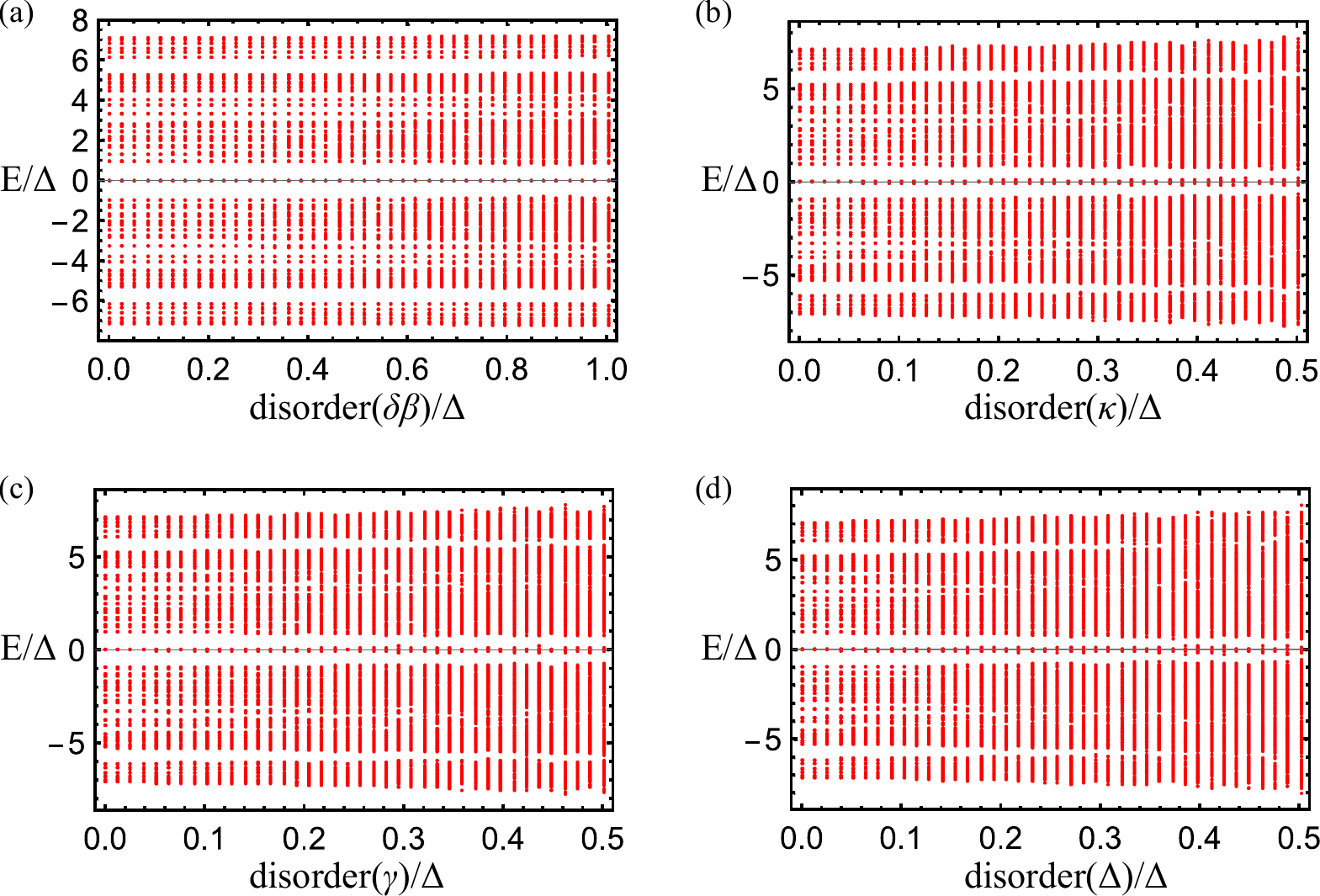}
    \caption{
    Evolution of finite-lattice spectra with disorder in (a) symmetric mode detuning $\delta \beta$, (b) $\kappa$, (c) $\gamma$, (d) $\Delta$. Parameters: $\kappa = \gamma = \Delta$, $14 \times 14$ site lattice. 
    }
    \label{evolution}
\end{figure*}

We examine the robustness of the zero-energy edge and corner states introducing several types of the disorder:
\begin{enumerate}
    \item Symmetric splitting $\delta\beta$ of the nearly degenerate $s$ and $p$ modes of a two-mode waveguide.
    \item Disorder in $s-s$ coupling $\kappa$.
    \item Disorder in $p-p$ coupling $-\gamma$.
    \item Disorder in $s-p$ coupling $\pm\Delta$.
\end{enumerate}
Numerical results presented in Fig.~\ref{evolution}(a-d) demonstrate that the energy of the zero modes is resilient to all above types of the disorder. This demonstrates the topological protection of the edge and corner states in the dipole-quadrupole topological insulator.

\bibliography{refs-supplement}